\documentclass[preprint,12pt]{elsarticle}

\usepackage{amssymb}
\usepackage{graphicx}
\usepackage{dcolumn}
\usepackage{bm}
\usepackage{amssymb}
\usepackage{epsfig}
\usepackage[utf8]{inputenc} 
\usepackage{ucs} 
\usepackage{amsmath} 
\usepackage{array}
\usepackage{float}
\usepackage{color}
\usepackage[usenames,dvipsnames]{xcolor}
\usepackage{epstopdf}
\usepackage{natbib}
\usepackage{lineno}
\usepackage{color}
\usepackage{subcaption}
\usepackage{caption}
\usepackage{hyperref}
\definecolor{red}{rgb}{1.0,0.0,0.0}
\definecolor{blue}{rgb}{0.0,0.0,1}
\newcommand{\T}{\tilde}

\newcommand{\be}{\begin{equation}}
\newcommand{\ee}{\end{equation}}
\newcommand{\bea}{\begin{eqnarray}}
\newcommand{\eea}{\end{eqnarray}}
\newcommand{\beas}{\begin{eqnarray*}}
\newcommand{\eeas}{\end{eqnarray*}}

\newcommand{\nn}{\nonumber}

\newcommand{\dpi}{(2\pi)}
\newcommand{\p}{\parallel}
\newcommand{\pp}{\perp}

\newcommand{\gmn}{g^{\mu\nu}}
\newcommand{\gma}{g^{\mu\alpha}}
\newcommand{\gan}{g^{\alpha\nu}}

\journal{Web of Conferences Journal}

\begin{document}
\begin{frontmatter}
\title{Prompt photon yield and $v_2$ coefficient from gluon fusion induced by magnetic field in heavy-ion collision}

\author{Jorge David Castaño-Yepes\corref{cor1}\fnref{a}}
\ead{jorge.castano@correo.nucleares.unam.mx}
\author[a,b]{Alejandro Ayala}
\author[b]{C. A. Dominguez}
\author[a]{L. A. Hernández}
\author[a]{Saúl Hernández-Ortíz}
\author[c]{María Elena Tejeda-Yeomans}
 \address[a]{Instituto de Ciencias Nucleares, Universidad Nacional Aut\'onoma de M\'exico, M\'exico Distrito Federal, C. P. 04510, M\'exico}
 
 \address[b]{Centre for Theoretical and Mathematical Physics, and Department of Physics, University of Cape Town, Rondebosch 7700, South Africa.}
 
 \address[c]{Departamento de Física, Universidad de Sonora, Boulevard Luis Encinas J. y Rosales, Colonia Centro, Hermosillo, Sonora 83000, México}
\begin{abstract}
We compute the production of prompt photons and the $v_2$ harmonic coefficient in relativistic heavy-ion collisions induced by gluon fusion in the presence of an intense magnetic field, during the early stages of the reaction. The calculations take into account several parameters which are relevant to the description of the experimental transverse momentum distribution, and elliptic flow for RHIC and LHC energies. The main imput is the strength of the magnetic field which varies in magnitude from 1 to 3 times the pion mass squared, and allows the gluon fusion that otherwise is forbidden in the absence of the field. The high gluon occupation number and the value of the saturation scale also play an important role in our calculation, as well as a flow velocity and geometrical factors. Our results support the idea that the origin of at least some of the photon excess observed in heavy-ion experiments may arise from magnetic field induced processes, and gives a good description of the experimental data.
\end{abstract}
\end{frontmatter}

%

%
\section{Introduction}
\label{intro}
It is well established that in heavy-ion experiments carried out at the CERN Large Hadron Collider and at the BNL Relativistic Heavy-Ion Collider, magnetic fields with intensities of several times the pion mass squared are produced, both in central and peripheral collisions \cite{intensity,Bzdak}. Studies on the centrality-dependence of these short-lived magnetic fields, show that their intensity along the reaction plane is small compared with  the intensity along the normal to the reaction plane.\\

The presence of a magnetic fields in a medium with high gluon occupation number \cite{McLerran} allows processes in which the photon production by gluon fusion can be achieved \cite{Ayala1,Ayala2}. This photon production mechanism together with the common sources calculated for syncroton radiation, bremsstrahlung, pair annihilation \cite{Zakharov, Tuchin} and modeled by hydrodynamical and transport calculations \cite{hydro-photons1,hydro-photons2, transport} could explain  the experimentally measured photon excess at low momentum in the invariant momentum distribution~\cite{experimentsyield, experimentsv2,experiments2}.\\

On the other hand, the precense of a magnetic field breaks the spatial isotropy in the photon emission which must have consequences for the $v_2$ harmonic coefficient or elliptic flow. This coefficient has been also calculated by hydrodynamical models and compared with ALICE and PHENIX measurements but its agreement is yet incomplete~\cite{review}.\\

 In this work we compute the photon production and $v_2$ harmonic coefficient from the fusion of low momentum gluons in the presence of a magnetic field. Our perturbative scheme is valid during times when the magnetic field reaches its maximum values and the shattered glasma is highly occupied by gluons that  can be described as quasiparticles \cite{quasiparticles}. These times are of order of $\tau_s\approx1/\Lambda_s$ or $\Delta\tau_s\simeq 1.5$ fm, with $\Lambda_s$ the saturation scale~\cite{Larry2, Lappi}. We explore a region of magnetic fields between 1 to 3 times the pion mass squared and we include a phenomenological expansion factor. 
 
\section{Photon Production by gluon fusion}
\label{sec-1}

\begin{figure}[H]
\begin{center}
\includegraphics[scale=.45]{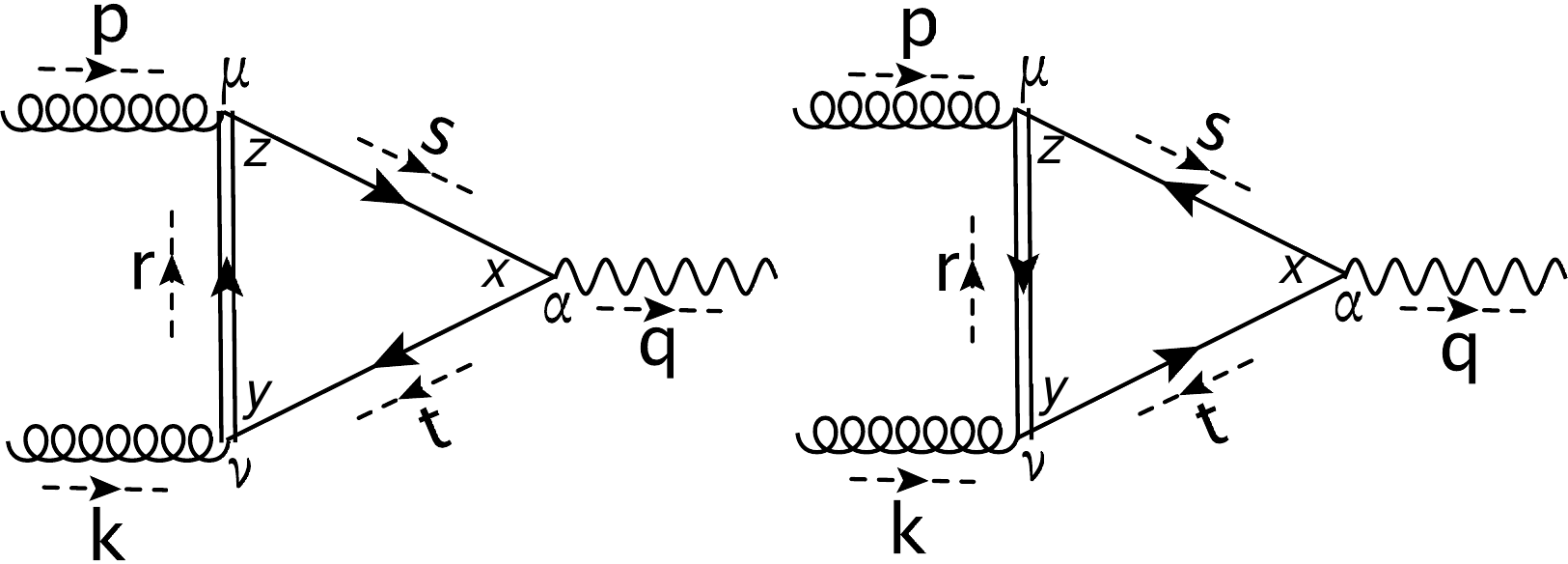}\includegraphics[scale=.45]{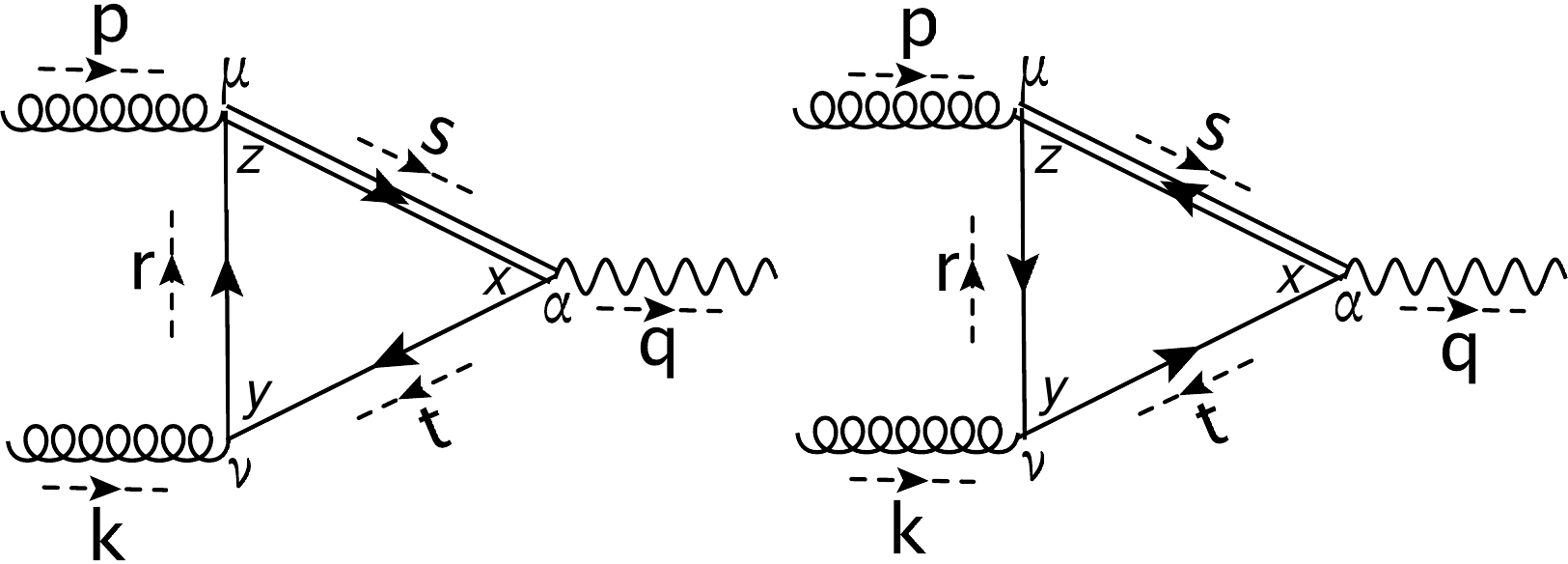}\\
\includegraphics[scale=.45]{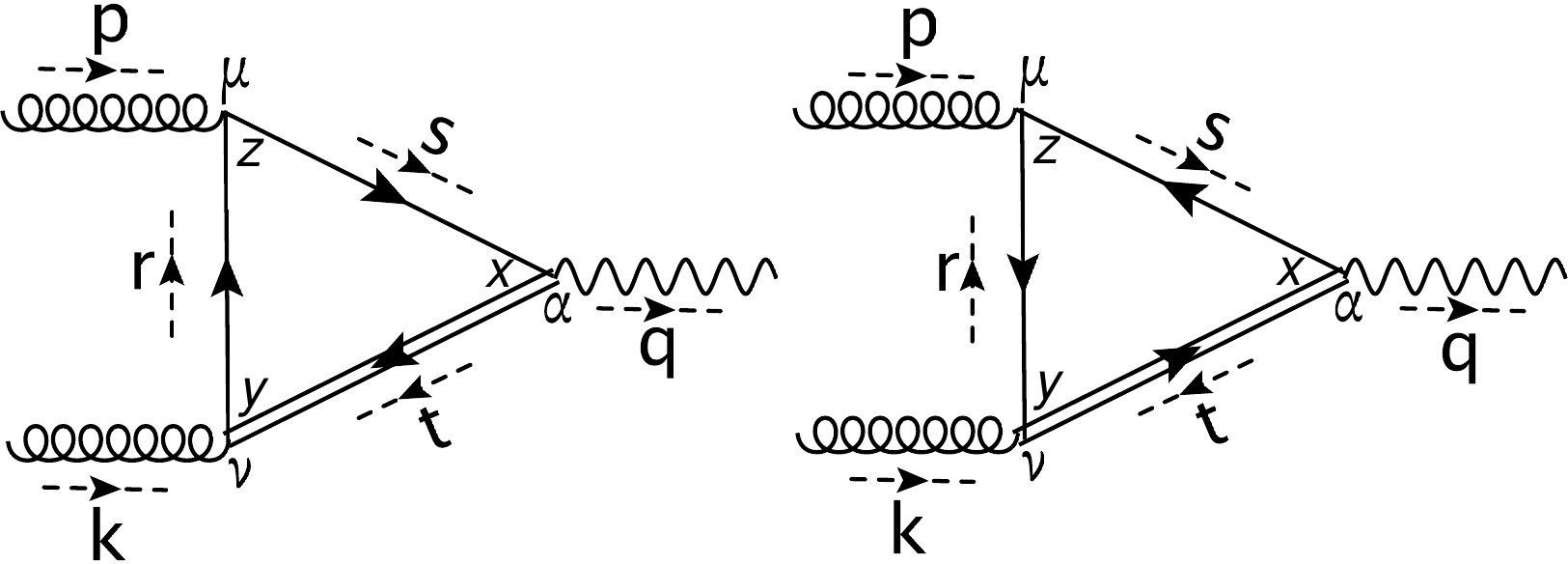}
\caption{Dominant contribution for photon production by gluon fusion in presence of a magnetic field. The double lines represent that the corresponding propagator is in the first Landau Level $S^{(1)}$. The single lines represents the propagator in the lowest Landau Level $S^{(0)}$. The arrows in the propagators represent the direction of the flow of charge. The arrows at the sides of the propagator lines represent the momentum direction.}
\label{fig1}
\vspace{-1cm}
\end{center}
\end{figure}

The sum of amplitudes of Fig. \ref{fig1} is given by
\bea
\mathcal{\T{M}}&=&-i\dpi^4\delta^{(4)}(q-k-p)\frac{|q_f|g^2\delta^{cd}e^{f(p_\pp,k_\pp)}}{32\pi\dpi^8}\nn\\
&\times&\left\{\left(\gma_\p-\frac{p^\mu_\p p^\alpha_\p}{p_\p^2}\right)h^\nu(a)-\left(\gmn_\p-\frac{p^\mu_\p p^\nu_\p}{p_\p^2}\right)h^\alpha(a)\right.\nn\\
&+&\left.\left(\gmn_\p-\frac{k^\mu_\p k^\nu_\p}{k_\p^2}\right)h^\alpha(b)-\left(\gan_\p-\frac{k^\alpha_\p k^\nu_\p}{k_\p^2}\right)h^\mu(b)\right.\nn\\
&+&\left.\left(\gan_\p-\frac{q^\alpha_\p q^\nu_\p}{q_\p^2}\right)h^\mu(c)-\left(\gma_\p-\frac{q^\mu_\p q^\alpha_\p}{q_\p^2}\right)h^\nu(c)\right\}\nn\\
&\times&\epsilon_\mu(\lambda_p)\epsilon_\nu(\lambda_k)\epsilon_\alpha(\lambda_q).
\eea
with $h^{\mu}(a)=(i/\pi)\epsilon_{ij}a^ig^{j\mu}_\pp$, $a_i=p_i+2k_i+i\epsilon_{im}p_m$;  $b_i=2p_i+k_i-i\epsilon_{im}k_m$, $c_i=k_i-p_i+i\epsilon_{im}(p_m+k_m)$ and
\bea
f\left(p_\pp,k_\pp\right)=\frac{1}{8|q_fB|}\left(p_m-k_m+i\epsilon_{mj}(p_j+k_j)\right)^2-\frac{1}{2|q_fB|}\left(p_m^2+k_m^2+2i\epsilon_{jm}p_mk_j\right).
\label{matrixelemafterapprox}
\eea

In order to compute Eq. (\ref{matrixelemafterapprox}) we considered one quark in the Lowest Landau Level ($S^{(0)}$) and two in the first excited Landau Level ($S^{(1)}$). Also we have been working in the massless quark aproximation and used the fact that the magnetic field is the dominan energy scale, i.e., $2|q_fB|\gg t_\p^2,\ s_\p^2,\ r_\p^2$.\\

Finally, when ignoring the magnetized medium dispersive properties, as in the present work, from energy-momentum conservation, the 4-vectors $p$ and $k$ need to be parallel to $q$: $   p^\mu=\left(\omega_p/\omega_q\right)q^\mu,\;k^\mu=\left(\omega_k/\omega_q\right)q^\mu$, therefore the invariant photon momentum distribution is thus given by
\bea
\omega_q\frac{dN^{\mbox{\tiny{mag}}}}{d^3q}&=&\frac{\chi {\mathcal{V}} \Delta \tau_s}{2(2\pi)^3}
\int\frac{d^3p}{\dpi^32\omega_p}\int\frac{d^3k}{\dpi^32\omega_k}
n(\omega_p)n(\omega_k)\nn\\
&\times&\dpi^4\delta^{(4)}\left(q-k-p\right)\frac{1}{4}\sum_{\mbox{\small{pol}},f}|{\mathcal{M}}|^2,\nn
\label{invdist}
\eea
where the high distribution number is given as in Refs.~\cite{Larry2, Krasnitz} and we sum over the three light flavors. The final result is shifted by the expansion  factor $\omega_{p,k}\to (p,k)\cdot u$. For simplicity we allow for a constant flow velocity $u^\mu=\gamma(1,\beta)$, with $\gamma=1/\sqrt{1-\beta^2}$. The coefficent $v_2$ results fom the Fourier decomposition of Eq. (\ref{invdist}) and from the weighed average
\bea
   v_2(\omega_q)=
   \frac{
   \frac{dN^{\mbox{\tiny{mag}}}}{d\omega_q}(\omega_q)\
   v_2^{\mbox{\tiny{mag}}}(\omega_q)
   +
   \frac{dN^{\mbox{\tiny{direct}}}}{d\omega_q}(\omega_q)\
   v_2^{\mbox{\tiny{direct}}}(\omega_q)} 
   {\frac{dN^{\mbox{\tiny{mag}}}}{d\omega_q}(\omega_q) 
   + 
   \frac{dN^{\mbox{\tiny{direct}}}}{d\omega_q}(\omega_q)}.
   \label{v2}
\eea

\section{Results and Discussion}
\label{sec-2}
\begin{figure}[h]
\begin{center}
\includegraphics[scale=.4]{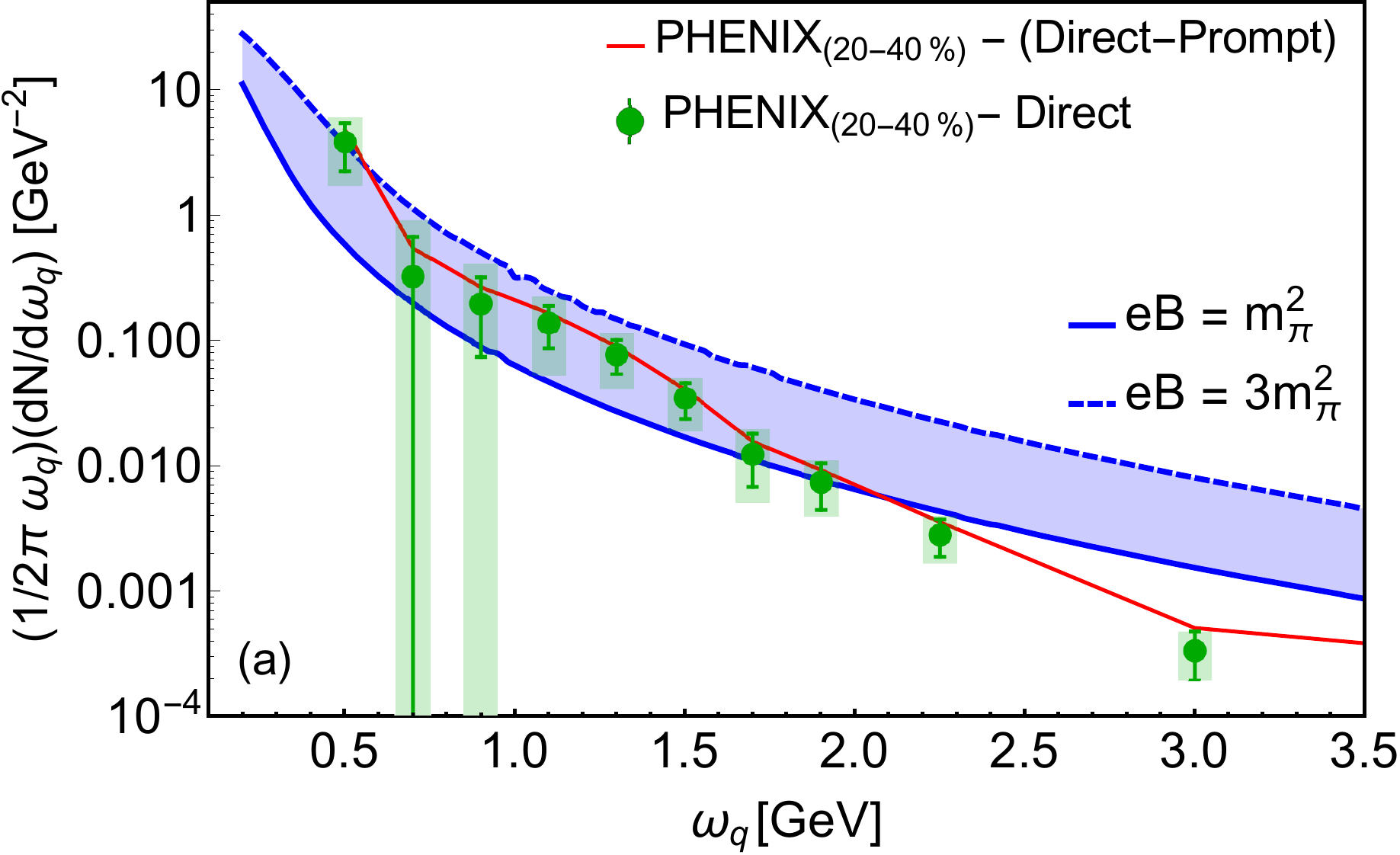}\includegraphics[scale=.4]{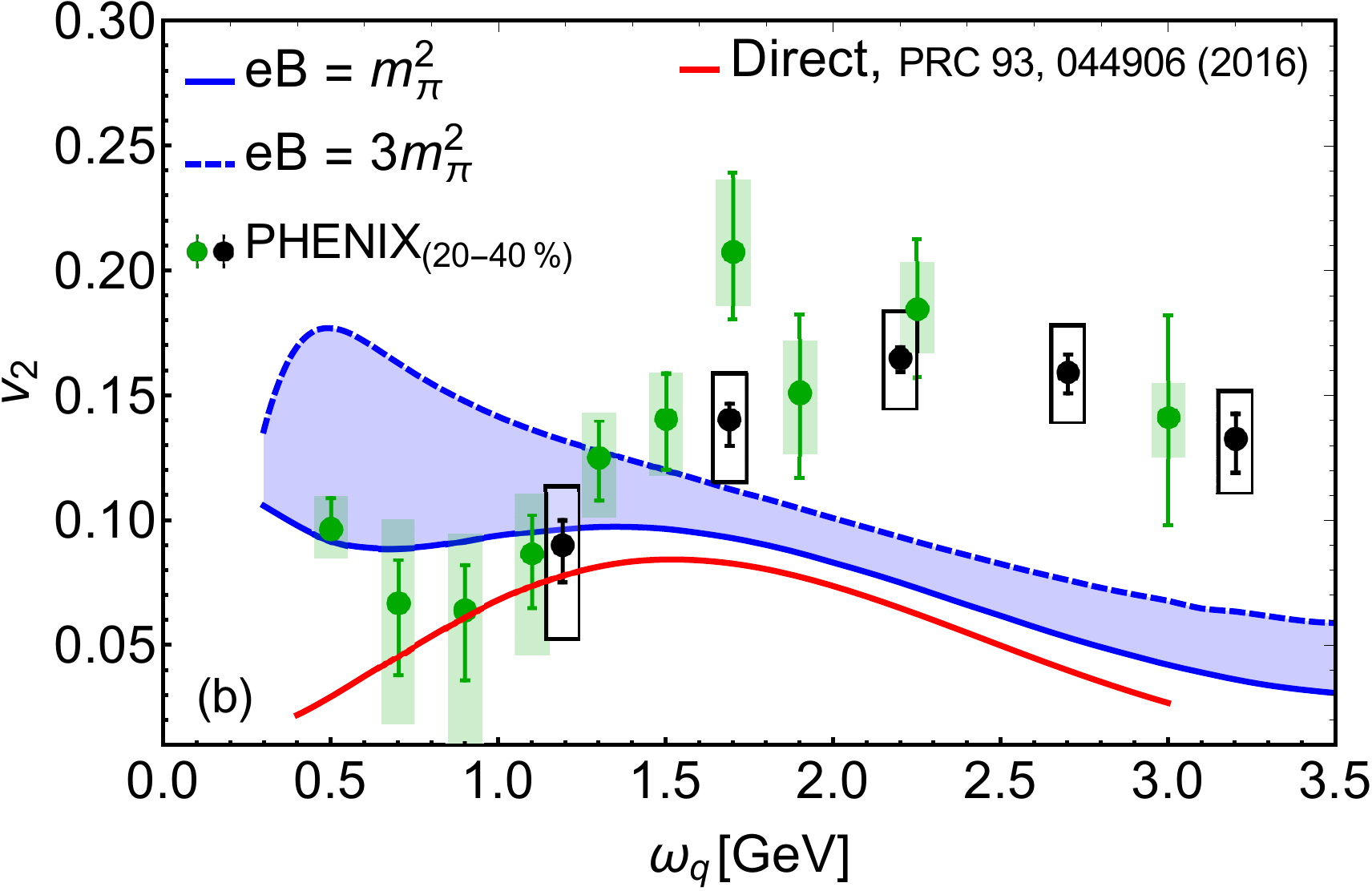}
\end{center}
\caption{(Left) Difference between PHENIX photon invariant momentum distribution~\cite{experimentsyield} and direct (points) or direct minus prompt (zigzag) photons from Ref.~\cite{hydro-photons1} compared to the yield from the present calculation. (Right) Harmonic coefficient $v_2$ combining the calculation of Ref.~\cite{hydro-photons1} and the present calculation compared to PHENIX data~\cite{experimentsv2}. Curves are shown as functions of the photon energy for central rapidity and the centrality range 20-40\%. Only the experimental error bars are shown. The bands show variations of the parameter $eB$ within the indicated ranges and computed with $\alpha_s=0.3$, $\Lambda_s=2$ GeV, $\eta=3$, $\Delta \tau_s=1.5$ fm, $R=7$ fm, $\beta=0.25$ and $\chi=0.8$.}
\label{fig2}
\end{figure}

Figure \ref{fig2}-(left) shows the results of Eq. (\ref{invdist}) compared with the diference between PHENIX~\cite{experimentsyield} data and the hydrodynamical calculations of Ref.~\cite{hydro-photons1}, wich are represented by the points. If the present calculation is considered as a yield, it provides a very good description of the excess of photons. Figure \ref{fig2}-(right) shows coefficient $v_2$ from Eq. (\ref{v2}) and compared with the direct photon result of Ref.~\cite{hydro-photons1} together with our calculation, also compared to PHENIX data~\cite{experimentsv2}. The curves are shown as functions of the photon energy for central rapidity and the centrality range 20$\%$ - $40\%$. We notice that the photon excess by gluon fusion helps to better describe the experimental data and highlights the importance of including the effects of magnetic fields in the early stages of the collision and its impact on the final state observables.

\end{document}